# Optical Imaging of Flavor Order in Flat Band Graphene

**Authors:** Tian Xie[1], Tobias M. Wolf[2], Siyuan Xu[1], Zhiyuan Cui[1], Richen Xiong[1], Yunbo Ou[3], Patrick Hays[3], Ludwig F Holleis[1], Yi Guo[1], Owen I Sheekey[1], Caitlin Patterson[1], Trevor Arp[1], Kenji Watanabe[4], Takashi Taniguchi[5], Seth Ariel Tongay[3], Andrea F Young[1], Allan H. MacDonald[2]\*, Chenhao Jin[1]\*

**Affiliations:**

[1]Department of Physics, University of California at Santa Barbara, Santa Barbara, CA, 93116, USA

[2]Department of Physics, University of Texas at Austin, Austin, TX, 78712, USA

[3] Materials Science and Engineering Program, School of Engineering for Matter, Transport, and Energy, Arizona State University, Tempe, Arizona 85287, USA

[4]Research Center for Electronic and Optical Materials, National Institute for Materials Science, 1-1 Namiki, Tsukuba 305-0044, Japan

[5]Research Center for Materials Nanoarchitectonics, National Institute for Materials Science, 1-1 Namiki, Tsukuba 305-0044, Japan

\* Corresponding author. Email: macdpc@physics.utexas.edu, jinchenhao@ucsb.edu

## Abstract

Spin- and valley flavor polarization plays a central role in the many-body physics of flat band graphene[1–21], with fermi surface reconstructions—often accompanied by quantized anomalous Hall[1–8] and superconducting state[9–14]—observed in a variety of experimental systems. Here we describe an optical technique that sensitively and selectively detects flavor textures via the exciton response of a proximal transition metal dichalcogenide layer. Through a systematic study of rhombohedral and rotationally faulted graphene bilayers and trilayers, we show that when the semiconducting dichalcogenide is in direct contact with the graphene, the exciton response is most sensitive to the large momentum rearrangement of the Fermi surface, providing information that is distinct from and complementary to electrical compressibility measurements. The wide-field imaging capability of optical probes allows us to obtain spatial maps of flavor orders with high throughput, and with broad temperature and device compatibility. Our work paves the way for optical probing and imaging of flavor orders in flat band graphene systems.

Flatband graphene systems are a versatile platform for engineering correlated and topological phenomena. While their phase diagrams vary remarkably with sample configuration parameters such as layer number and relative alignment[1–21], a feature common to all systems is that small changes in the carrier density and other experimental tuning parameters drive flavor order transitions (FTs) where the relative occupation of the (nominally degenerate) electron orbitals with differing spin and valley polarization changes. In both crystalline and twisted graphene systems, these transitions are often accompanied by superconducting domes, suggesting that flavor symmetry breaking may play an important role in superconducting pairing[9–11,22–24]. To refine the understanding of the phase diagram, FTs have been investigated by various experimental techniques including electrical transport[7–9,15–17], measurements of the thermodynamic compressibility and magnetization[7,9,17,25,26], and scanning tunneling microscopy[15,27–29]. However, these techniques all come with drawbacks: bulk electrical measurements typically fail in the face of spatial inhomogeneity and become less expressive in superconducting states, while scanning tunneling measurements are incompatible with the common dual-gated geometry. Moreover, all measurements are inherently low bandwidth, precluding studies of dynamics.

Here we describe an optical technique that addresses some of these challenges. Fig. 1a illustrates the device scheme, in which a $WSe_2$ sensor layer is placed in direct contact with a target graphene system. The short-range interaction between graphene and $WSe_2$ leads to a shift in the quasi-particle bandgap of the $WSe_2$ that depends on the flavor polarization of the graphene layer; this can be read out optically via the reflection contrast (RC) spectra. As we detailed below, our measurement configuration provides information that is distinct from electrical compressibility measurements or exciton sensing using a physically separated $WSe_2$ layer, which both detect the zero-wavevector limit of the polarzibility[30,31], and so offers a novel capability in identifying flavor transitions.

**Optical sensing of FT**

We first study a rhombohedral trilayer graphene (RTG) device D1. Fig. 1b shows the inverse compressibility of the device measured at 3K (see methods). The phase diagram features spontaneous formation of flavor orders on both electron and hole doping sides, consistent with previous reports[7,10]. Owing to sample inhomogeneity, two sets of patterns can be observed that are offset from each other (see Extended Data Fig. 1). Fig. 1c shows RC of the same sample measured at displacement field $D = 0.673V/nm$ (dashed line in Fig. 1b). We focus on the spectral range near the 2s exciton resonance of $WSe_2$ (see Extended Data Fig. 2 for full spectra). The 2s exciton energy shows two prominent kinks on the hole doping side, reminiscent of the two low compressibility lobes in Fig. 1b. The lower panel in Fig. 1c compares the fitted 2s exciton energy and the inverse compressibility (see methods). The optical spectrum reproduces all features of the compressibility on the hole doping side, while the response on the electron side is rather weak. Such asymmetry can be naturally understood from a short-range interaction between $WSe_2$ and graphene. Under the large displacement field applied, doped holes and electron primarily reside the top and bottom graphene layers, respectively. The much weaker responses from $WSe_2$ to electrons than holes indicate that $WSe_2$ primarily interacts with charges in the top (closest) layer with an interaction range $\Delta r < 1nm$. Our sensing scheme therefore also provides a sensitive probe of layer polarization. Fig. 1d summarizes the 2s exciton energy over the same parameter range as Fig. 1b (see methods). The optical phase diagram matches well with the electrical one, except that features appear only in the top left and bottom right quadrants owing to the layer polarization sensitivity.

Having demonstrated the ability to detect FT, we now show that our technique provides unique information. Figure 1e compares electrical capacitance and optical RC measurements under the same experimental condition of $D = 0V/nm$ and $B_z = 3T$ (see Ext Data Fig. 1 for more results). A series of incompressible peaks emerge in capacitance, corresponding to gaps between Landau levels. Surprisingly, none of them appear in RC spectra. In capacitance the incompressible peaks are quite prominent, several times larger than the FT-induced features (Fig. 1c). If the optical response were effectively measuring compressibility one would expect, in contradiction to our observations, similarly strong features.

To gain further insight, we apply our sensing technique to alternating-twisted-magic-angle-trilayer graphene (MATTG). Fig. 2d shows the four-probe longitudinal resistance of MATTG device D2 with twist angle of 1.43 degree (see methods). The phase diagram is qualitatively consistent with previous reports[11,14,15,32] in that resistive states emerge at integer moiré fillings from $v = 0$ to 4 under large displacement field ($v = 1$ corresponds to one electron per moiré period). At smaller displacement fields, the resistive behaviors at integer fillings become weaker, suggesting Fermi surface resets instead of gaps[11,15]. Interestingly, optical measurement of the same device shows the opposite trend. At zero displacement field (Fig. 2b), the 2s exciton resonance shows prominent cascade features at integer fillings, which becomes weaker at larger displacement fields (Fig. 2, a and c). See Extended Data Fig. 3 for more data. Fig. 2e summarizes the 2s exciton energy across the entire phase diagram. While the emergence of features around integer fillings is consistent with transport measurement, their displacement field dependencies are in sharp contrast. Optical sensing does not weigh the gaps at large displacement field heavily but is sensitive to FT-induced Fermi surface reconstructions.

We have also performed measurements on Bernal bilayer graphene (BBG) in the quantum Hall regime. The upper panel in Fig. 3a shows the RC spectra of BBG device D3 under $B_z = 3T$ and zero displacement field. The lower panel shows a comparison between the 2s exciton energy (black) and incompressibility (red) under the same measurement conditions. See Extended Data Fig. 4 and 5 for more results. A series of chemical potential jumps, observed as peaks in incompressibility, appear at even filling factors $v \in (-4,4)$ and at higher filling factors only at the cyclotron gap filling factors, which are multiples of 4 because of spin-valley degeneracy. The peaks within $v \in (-4,4)$ are related to flavor ferromagnetism. The optical measurement again shows quite distinct behavior. Instead of having features at even filling factors, the 2s exciton energy oscillates rapidly in the $v \in (-4,4)$ interval between minima at odd filling factors and maxima at even filling factors. No strong features are seen at higher filling factors, even when the Fermi level lies in a cyclotron gap. The lack of gap features at high filling factors is consistent with our observations in RTG (Fig.1e) and MATTG (Fig. 2e), and this makes the prominent features at $\nu \in (-4,4)$ even more surprising.

**Selective detection of FT**

Our investigations across multiple flat band graphene systems indicate that the optical sensing technique here provides qualitatively different information from electrical measurements and has unique FT sensitivity. Our results also contrast with those from the common exciton-sensing configuration with an hBN spacer, which largely reproduces electrical compressibility measurements, e.g. in the detection of graphene Landau levels[31]. To elucidate the origin of FT sensitivity, it is helpful to examine the role of an hBN spacer. To this end, we compare the RC spectra of device D3 and another BBG device D4 with a WSe$_2$ sensor layer and a 5nm hBN spacer, as shown in Fig. 3b and 3c. Under similar measurement conditions, we observe two major differences. First, the 2s exciton without a hBN spacer appears at a much lower energy (Fig. 3b), indicating much stronger interaction between WSe$_2$ and graphene. Second, at large displacement field, the exciton responses without a hBN spacer show prominent asymmetry (Fig. 3b) between electron and hole doping while those with a spacer remain largely symmetric (Fig. 3c). The lack of layer sensitivity in the latter case suggests that the interaction is long-range in nature, which also explains the much weaker interaction strength. We therefore conclude that the WSe$_2$-graphene interactions without (with) an hBN spacer are dominated by strong (weak) short- (long-)range interactions.

We have formulated a quantitative theoretical interpretation of our adjacent layer exciton sensing, one that also sheds light on the distinct information supplied by exciton sensing with hBN spacers. Our starting point is the successful GW theory of excitons[33,34] in TMDs (the "sensing layer"), within which the influence of a nearby 2D material (the "target layer") with negligible hybridization is captured exactly by adding a screening correction to Coulomb interactions $V_C \to V_C + \chi V_D^2$. Here $\chi$ is the target layer density response function, $V_D = 2\pi e^2 e^{-qd}/q$ is the interlayer Coulomb interaction and $d$ is the layer separation. $\chi V_D^2$ captures the contribution to the interaction between two electrons in the sensing layer that is mediated by charge density response in the target layer. Due mainly to reduced dimensionality, The 2s exciton has a rather small binding energy that is insensitive in absolute terms to screening[35–38]. Its resonance energy is then mainly determined by the quasi-particle bandgap of WSe$_2$. Because the carrier-density dependent part of the target layer response is at long wavelengths compared to a graphene lattice constant, the quasiparticle bandgap change reduces to a valence band exchange correction (see Supplementary Information). When the GW approximation is used for $\chi$, the quasiparticle band gap $E_{gap}$ is given by

$$E_{\text{gap}} \to E_0 + \int \frac{d^2q}{(2\pi)^2} \frac{\Pi_{22}\left(\boldsymbol{q}; \omega = \frac{\hbar^2 q^2}{2m^*}\right) V_D(q)^2}{1 - V_S(\boldsymbol{q})\Pi_{22}\left(\boldsymbol{q}; \omega = \frac{\hbar^2 q^2}{2m^*}\right)} \quad (1)$$

where $E_0$ is the quasi-particle bandgap of bare WSe$_2$, $\Pi_{22}$ is the non-interacting target layer response function, $m^*$ is the WSe$_2$ valence band effective mass and $V_S = 2\pi e^2/q$. An hBN spacer increases $d$ and decreases the momentum cutoff in the integral to $q_c < 1/d$. Fig. 3d illustrates such momentum cutoff implied by $V_D$ for three representative interlayer distance that corresponds to the adjacent graphene layer ($d = 2a_G$), the distant graphene layer in RTG ($d = 5a_G$), and the case with a 5nm hBN spacer ($d = 20a_G$), respectively. The small range of relevant momentum in the spacer case (green) captures the property that large-$q$ charge fluctuations do not produce a significant electrical potential in a distant layer. In the limit of thick hBN spacer and large $d$, $q_c \to 0$. $E_{\text{gap}}$ then depends on the graphene polarizability in the long wavelength and static limit, which is directly related to its compressibility. Therefore, the exciton sensing scheme with an hBN spacer largely reproduces results from electrical measurements.

The case without an hBN spacer is distinctively different since $d \sim 0.5\,nm$ for the closest graphene layer. $E_{\text{gap}}$ senses changes in the large-$q$ parts of graphene polarizability, which typically dominate due to their larger phase space (Fig. 3d). Our sensing scheme therefore mainly probes the large-$q$ polarizability of graphene, which is inaccessible to electrical measurements. The observed layer sensitivity (Fig. 1) is a direct manifestation, where the bandgap shifts induced by the adjacent and distant graphene layers in RTG differ by several times. As illustrated in Fig. 3e, the difference between the two cases (blue and red) only becomes prominent at large-$q$. The much larger bandgap shift from the adjacent layer confirms the dominance of large-$q$ contribution. This unique capability allows it to identify physics unrelated to the appearance of charge gaps, such as a cyclotron gap, that mainly affects the small-$q$ part of polarizability (red dashed line). It also explains its sensitivity to FT since FT involves reconstruction of the entire Fermi surface and modifies the large-$q$ polarizability up to several times of $k_F$ (blue dashed line). In the supplementary material we show that the 2s exciton energy changes that accompany FT in RTG (Fig. 1) agree quantitatively with the calculations.

The surprising exciton energy oscillations we have discovered in the small filling factor $\nu \in (-4,4)$ regime of BBG provide another excellent example of the new capabilities. As illustrated in Fig.3e, we interpret the minima in the 2s exciton energy at odd filling factors as evidence for orbital-polarized states with differential occupation between the *n*=0 and *n*=1 orbitals, which lead to strong screening over a wide range of wavevectors from inter-orbital contribution (see Supplementary Information); and the maxima at even filling factors as evidence for states in which both orbitals of a given flavor are completely occupied or empty. The differences between even and odd fillings only appear at nonzero $q$ (Fig. 3d), therefore the oscillation shows up in optical sensing but not compressibility (Fig. 3a). The convenient optical probe of the orbital content of fractional states in the $\nu \in (-4,4)$ regime of BBG could aid efforts to optimize robust non-Abelian quantum Hall states in the bilayer graphene platform[39–41].

**Wide field imaging of FT**

Besides FT sensitivity, our technique also offers wide-field imaging capability to capture spatial patterns of FT with high throughput. Fig. 4a, b shows the optical microscope image and reflection contrast spectra of a magic angle twisted bilayer graphene (MATBG) device D5. FT has been widely reported in MATBG, giving rise to Dirac revivals and Chern insulators at integer fillings[1,26,42–47]. Indeed, we observe clear features in 2s exciton resonance at integer moiré fillings $v = \pm 1$ to $\pm 4$ (orange arrows)[48,49]. On the other hand, MATBG is known for its extreme sensitivity to twist angle and intrinsic spatial inhomogeneity from lattice relaxation. Fig. 4c shows reflection contrast on a different spot in the same device, where we only observe the band insulator at $v = \pm 4$ but no features in between. In transport measurement of this device (Extended Data Figure. 6), we consistently observe strongly insulating states at $v = \pm 4$, while the features at $\pm 1$ to $\pm 3$ are generally weak and inconsistent between different source drain configurations. These observations exemplify a common challenge plaguing the study of twisted graphene systems, where devices vary strongly and it can often be difficult to extract intrinsic physics[3,26]. For example, different transport phenomena can be dominated by different conducting channels and may not be directly correlated.

Our technique offers a potential solution. As a demonstration, we perform wide-fielding imaging of the $v = 4$ band insulator and $v = 2$ cascade feature in Fig. 4d and 4e, respectively. This allows us to extract a spatial map of twist angle from the charge density at $v = 4$, and a map of the correlation strength from the prominence of cascade feature at $v = 2$. Each map is obtained in 15 minutes without spatial scanning (see methods and Extended Data Fig. 7). The cascade features only appear in a small spatial region close to the left edge of this device, which explains the weak and inconsistent features in transport. By comparing the FT map and angle map, we find that the cascade features emerge in a twist angle range between 1.01 and 1.07 degrees and is the most prominent at angle around 1.04 degree.

The high-throughput imaging capability of our technique is further augmented by its broad environment compatibility. It encodes low-energy flavor physics into exciton responses at much higher energy scale and is less susceptible to noises. Fig. 4f shows the 2s exciton energy at different temperatures up to 50K. Excitons resonances remain largely unchanged over this temperature range, allowing us to directly track melting of the cascade features. Interestingly, the cascade features remain visible at 50K, consistent with previous reports from chemical potential measurements[25,26,50,51] and is an order an order of magnitude higher than the temperature at which hysteresis of isospin ferromagnetism disappears[1,4,42,43]. This may suggest the existence of vestigial FT or flavor fluctuations over a broad temperature range[25,50].

**Discussions and outlook**

The reported technique opens several exciting opportunities in studying FT in flatband graphene systems and their interplay with other correlated phases. Its FT sensitivity offers an attractive approach to disentangle flavor orders and fluctuations from ordinary gaps and Fermi surface distortions such as nematicity[52–56], thereby shedding new light on the roles of these instabilities. The high-throughput imaging capability, along with the wide temperature and device geometry compatibility, enables investigation of FT spatial patterns near and across critical points. A particularly exciting opportunity lies in in-situ imaging of FT throughout the superconductivity domes in magic angle multi-layer graphene[22,23], which can be correlated to transport measurements to disentangle extrinsic and intrinsic effects and potentially elucidate the interplay between FT and superconductivity. By establishing an optical technique to detect FT, our work also paves the way for dynamic manipulation and investigation of flatband graphene systems using ultrafast light pulses, such as Floquet engineering of FT and studying its non-equilibrium dynamics.

## Methods

### Sample fabrication

The preparation of multilayer graphene, hexagonal boron nitride (hBN), and tungsten diselenide ($WSe_2$) flakes involves mechanical exfoliation of bulk crystals onto silicon substrates with a 285 nm silicon oxide layer. Rhombohedral domains within trilayer graphene flakes are identified using a Horiba T64000 Raman spectrometer equipped with a 488-nm mixed-gas Ar/Kr ion laser beam. Subsequent isolation of the rhombohedral domains is performed utilizing a Dimension Icon 3100 atomic force microscope[57,58].

All Van der Waals heterostructures are constructed through a standard dry-transfer technique employing a poly (bisphenol A carbonate) (PC) film on a polydimethylsiloxane (PDMS) stamp. The fabrication process involves initially creating the lower hBN/graphite part, releasing them onto a 90nm $Si/SiO_2$ substrate. The removal of polycarbonate residue on the sample is accomplished by dissolving it in chloroform, followed by rinsing with isopropyl alcohol and annealing at 375 °C. The upper part of the heterostructure is separately assembled and transferred onto the lower part. This stacking sequence is meticulously implemented to minimize mechanical stretching of the multilayer graphene. Standard electron-beam lithography, dry-etching processes, and vacuum deposition are employed to fabricate electrodes for electrical contacts (~150 nm gold with ~5 nm chromium and ~15nm palladium adhesion layers).

### Calibration of carrier density, displacement field, and twist angle

Carrier densities in all devices are calibrated from the hBN thickness measured by a Dimension Icon 3100 atomic force microscope. Using hBN dielectric constant $\varepsilon_{hBN}$=3.52, we compute the geometrical capacitance per unit area $c_{t,b}=\varepsilon_{hBN}\varepsilon_0/d_{t,b}$ between the top/bottom gate and sample, where $d_t$ ($d_b$) is the top (bottom) hBN thickness. The charge density and displacement field are obtained as $n_0 = (c_tV_t +c_bV_b)/2e$ and $D = (c_tV_t - c_bV_b)/2\varepsilon_0$, respectively, where $V_t$ ($V_b$) is the top (bottom) gate voltage and $e$ is elementary charge.

The twist angle of MATBG and MATTG are extracted from the cascade features at superlattice filling factors $v=\pm4$ (Fig. 2 and Fig. 4). From the corresponding carrier density $n_{v=4}$, the twist angle $\theta$ was obtained from $n_{v=4} = (8\theta^2)/(\sqrt{3}a_0^2)$, $a_0$=0.246nm is the graphene lattice constant.

### Reflection contrast (RC) measurement

The devices were mounted in a closed-cycle cryostat (Quantum Design, OptiCool) for all optical experiments with a base temperature of 3 K. A broadband tungsten lamp was beam-shaped by a single mode fiber and subsequently collimated by a lens. The light was focused onto the sample by an objective (NA=0.45), resulting in a beam diameter of approximately 1μm on sample with a power of approximately 20nW. The reflected light was collected by a liquid-nitrogen-cooled CCD camera coupled with a spectrometer. The reflection contrast was computed as RC=(R'−R)/R, where R' and R represent the reflected light intensity from regions with and without the sample, respectively. Keithley 2400 source meters were employed to apply gate voltages to adjust the charge density.

**Extraction of 2s exciton energy**

We extracted the 2s exciton energy at each carrier density from the local maximum in the slope of RC vs. probe energy (Extended Data Fig. 2a and 2c). The obtained 2s exciton energy shows a smooth decreasing background with increasing charge density due to stronger screening. Such background dominates the exciton energy shift in MATTG owing to the large range of carrier density. To highlight the cascade features associated with the FT, we fitted the smooth background for hole (electron) side using a $3^{rd}$ ($7^{th}$)-order polynomial (Extended Data Fig. 2d, Orange curve). The background-subtracted 2s exciton energy shows clear cascade features at integer fillings (Extended Data Fig. 2d). The same background was used for all displacement fields to ensure that no artifacts were introduced (Fig. 2e).

**Capacitance and transport measurement**

Penetration field capacitance measurements were performed on $WSe_2$/RTG device D1 and $WSe_2$/BBG graphene devices D3. The device capacitance $c_p$ was isolated from the environment using a low-temperature capacitance bridge[59]. The inverse compressibility $\kappa$ was obtained from $c_p$ through $c_p = c_t c_b/(c_t+c_b+\kappa^{-1}) \approx \kappa c_t c_b$[60]. The magnitude of $\kappa$ increases when the sample is incompressible (gapped) and decreases when it is compressible (conducting). The measurement of $\kappa$ involved applying a fixed AC excitation (17–88 kHz) to the top gate. The phase and amplitude of a second AC excitation of the same frequency were adjusted and applied to a standard reference capacitor ($c_{ref}$) on the low-temperature amplifier to balance the capacitance bridge. A commercial high-electron-mobility transistor (FHX35X) transformed the small sample impedance to a 1 kΩ output impedance, yielding a gain of about 1,000. The DC components of $V_t$ and $V_b$ were supplied by Keithley 2400 source meters and were connected to the corresponding gate though bias tee. Additional electrodes were patterned in $WSe_2$/MATTG device D2 and $WSe_2$/MATBG device D5 for electrical transport. Four-point longitudinal resistance was obtained by supplying an AC current of 10nA amplitude at frequency of 17.777 Hz.

**Widefield imaging of cascade features**

A broadband supercontinuum laser (YSL photonics SC-OEM) was filtered by a home-built double monochromator to generate probe light of tunable center wavelength and <0.2 nm full width at half maximum (FWHM). The probe light was expanded before focusing on the sample, giving rise to a field of view of ~900μm² that covers the entire device. The wide-field image of sample was collected by an EMCCD camera (ProEM-HS 512BX3) without spatial scanning. To obtain a map of the cascade features, we tuned the probe light energy to be slightly above the $WSe_2$ 2s exciton resonance and took a sample reflection image at each carrier density. Ordinarily, the 2s exciton energy redshifts with increasing carrier density, leading to decrease of sample reflection at the probe energy. On the other hand, the cascade features at integer fillings lead to abnormal blueshifts of 2s exciton energy with increasing carrier density (Fig. 4b) and thereby increase of sample reflection. This allowed us to extract both carrier density and strength of the cascade features by comparing sample images at neighboring carrier density.

We further employed a lock-in algorithm to improve the signal to noise ratio. The carrier density in the device was modulated at 66Hz by a small AC gate voltage $\Delta V_g = 0.01$V on top of the DC gate voltage $V_g$. The EMCCD camera was externally triggered and synchronized with the AC gate modulation, thereby directly obtaining the differential reflection image of the sample between slightly different carrier densities. Extended video 1 and 2 show the obtained differential reflection images for a range of carrier densities near $v = 2$ and $v = 4$ of MATBG, respectively. Extended Data Fig. 7 shows the carrier density-dependent differential reflection near $v = 4$ for a representative spatial spot (blue boxed pixel). The non-monotonic dip from the cascade features was fitted by an 2nd-order polynomial, from which we extracted the carrier density and the amplitude of the $v = 4$ cascade feature. Similar fitting was performed on each pixel for carrier densities near $v = 4$ and $v = 2$, from which we obtained a map of the twist angle and correlation strength (Fig. 4d and e).

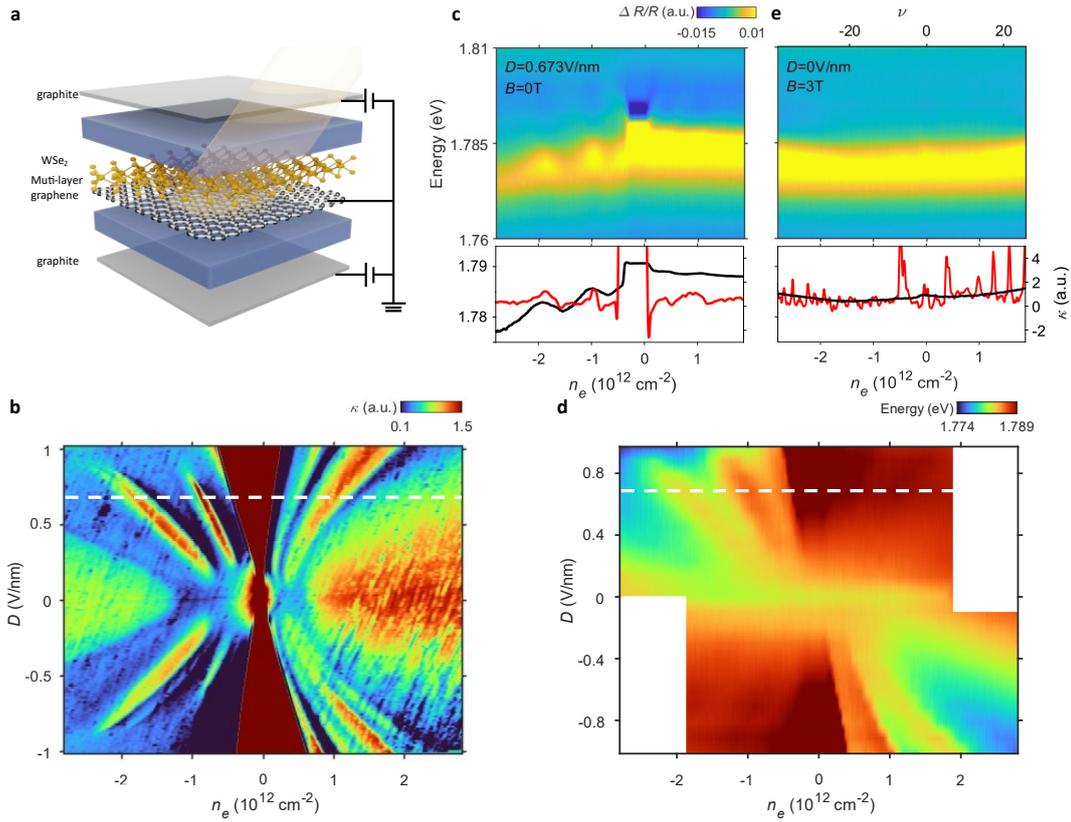

**Figure 1. Optical sensing of FT in RTG**. (**a**) Schematics of device configuration. A WSe$_2$ sensing layer is placed adjacent to flatband graphene without a spacer. The short-range interaction between graphene and WSe$_2$ imprints flavor orders of graphene into WSe$_2$ exciton responses. (**b**) Displacement-field and carrier-density dependent incompressibility of RTG device D1. (**c**) Upper panel: RC of device D1 at $D$=0.673V/nm and $B$=0T (white dotted line in (**b**) and (**d**)) near WSe$_2$ 2s exciton resonance. Lower panel: extracted 2s exciton energy (black) and comparison to incompressibility (red). The exciton energy shift fully captures FT on the hole side. (**d**) Displacement-field and carrier-density dependent 2s exciton energy of device D1. Features are only observed in the top-left and bottom-right quadrant owing to the sensitivity to layer polarization. (**e**) Upper panel: RC of device D1 at $D$=0V/nm and $B$=3T. Lower panel: extracted 2s exciton energy (black) and comparison to incompressibility (red). The prominent incomparability peaks from charge gaps do not show up in optical sensing, in contrast to the FT in (**c**). All measurements are performed at a temperature of 3K.

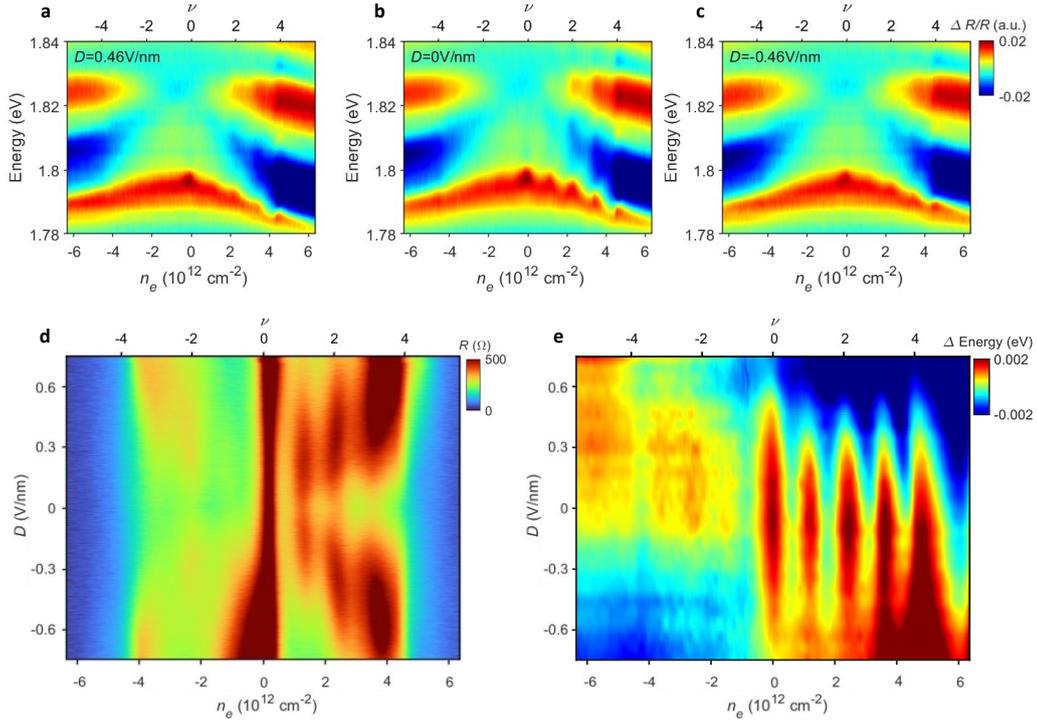

**Figure 2. Selective sensing of FT**. (**a**)-(**c**) RC of MATTG device D2 at *D*= (**a**) 0.46, (**b**) 0, and (**c**) -0.46V/nm. The 2s exciton resonance shows cascade features at integer fillings, which becomes weaker at larger displacement field. (**d**)(**e**) Longitudinal resistance (**d**) and 2s exciton energy (**e**) of device D2 as a function of displacement field and carrier density. The insulating features at integer fillings in transport measurements become more prominent at larger displacement field due to the transition from Fermi surface resets to charge gaps. In contrast, the optical sensing is more sensitive to FT-induced Fermi surface reconstruction at low displacement field than the charge gaps at high displacement field. All measurements are performed at a temperature of 3K.

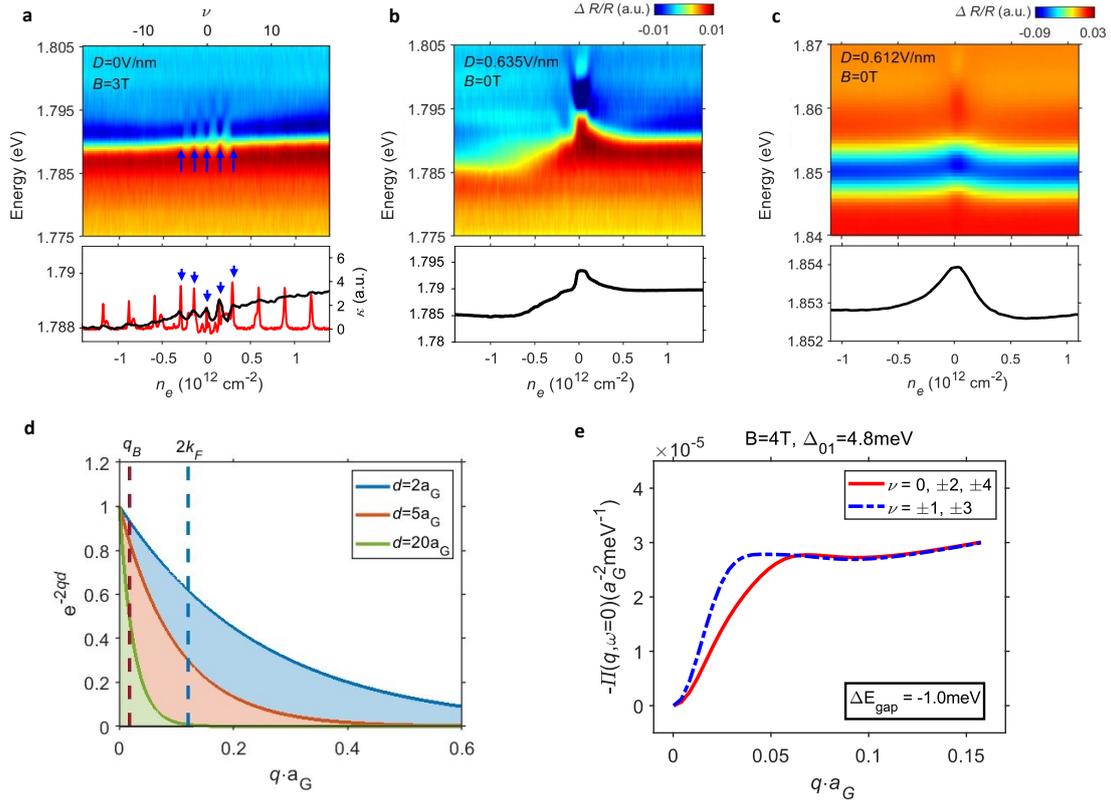

**Figure 3. Probing orbital polarization in BBG**. (**a**) Upper panel: RC of BBG device D3 at *D*=0 and B=3T. Lower panel: extracted 2s exciton energy (black) and comparison with incompressibility (red). The strong incompressibility peaks from cyclotron gaps do not show up in RC. Instead, an oscillation of 2s exciton energy is observed between even and odd fillings within the zeroth Landau level. Blue arrows mark even fillings within the octet of the zeroth Landau level. (**b**)(**c**) Comparison between device D3 without an hBN spacer (**c**) and BBG device D4 with a ~5nm hBN spacer (**e**) under similar measurement configurations. Their distinct behaviors indicate the dominance of short-range and long-range interactions, respectively, as detailed in the text. (**d**) Momentum-cutoff for three representative interlayer distances *d*. $a_G$ = 0.246nm is the graphene lattice constant. Vertical dashed lines mark the momentum range of polarizability change from a cyclotron gap at *B*=3T (inverse magnetic length $q_B$, red) and from a representative FT in RTG (Fermi momentum $k_F$, blue). (**e**) Calculated static polarizability $\Pi(q,\omega=0)$ of graphene at even (red) and odd (blue) Landau level fillings under magnetic field *B*=4T. The finite-*q* part of graphene polarizability is enhanced at odd filling factors when *n*=0 and *n*=1 orbitals are alternately occupied, which can be uniquely accessed in adjacent layer exciton sensing as an energy shift $\Delta E_{gap}$. $\Pi(q,\omega) \approx \Pi(q,0)$ for $\omega \ll \Delta_{01}$ =4.8 meV, where $\Delta_{01}$ is the orbital splitting.

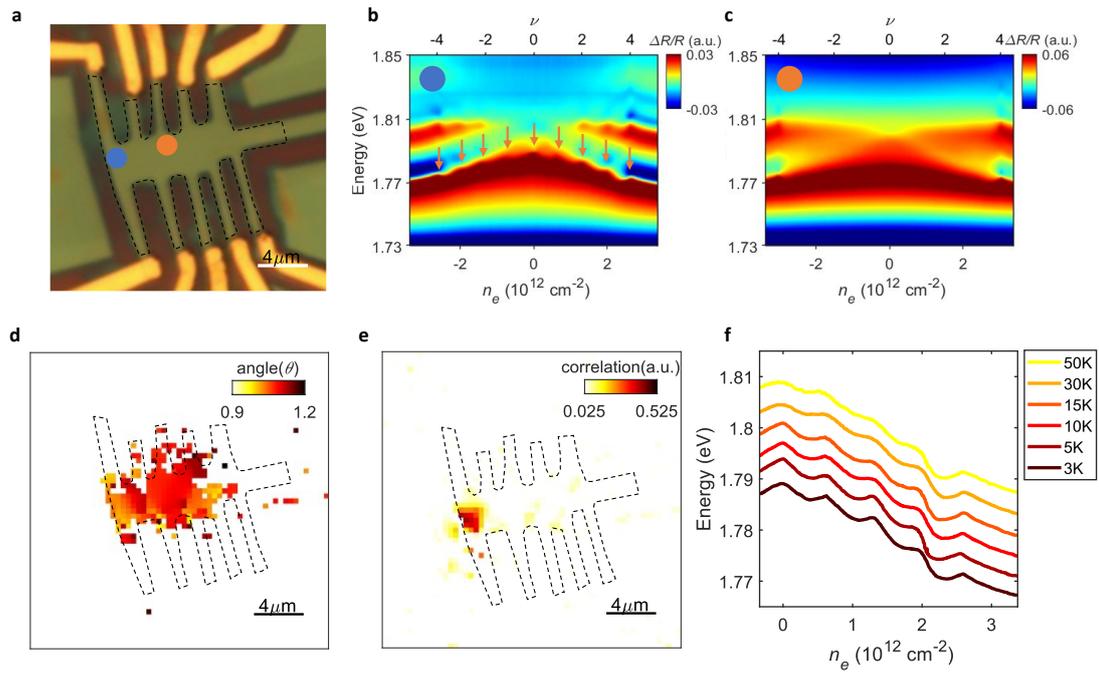

**Figure 4. Wide field imaging of FT**. (**a**) Optical microscope image for MATBG device D5. Scale bar: 4μm. (**b**)(**c**) RC of representative magic-angle (**b**) and non-magic angle (**c**) spots in device D5 with local twist angle of 1.04° and 1.14°, respectively. Their locations are marked by blue and orange dots in (**a**). (**d**)(**e**) Spatial map of twist angle (**d**) and correlation strength (**e**) extracted from the $v$=4 and $v$=2 cascade features, respectively. Both maps are obtained by wide-field imaging without scanning. (**f**) Temperature dependence of the extracted 2s exciton energy. The cascade features persist to above 50K.

# Extended Data Figures

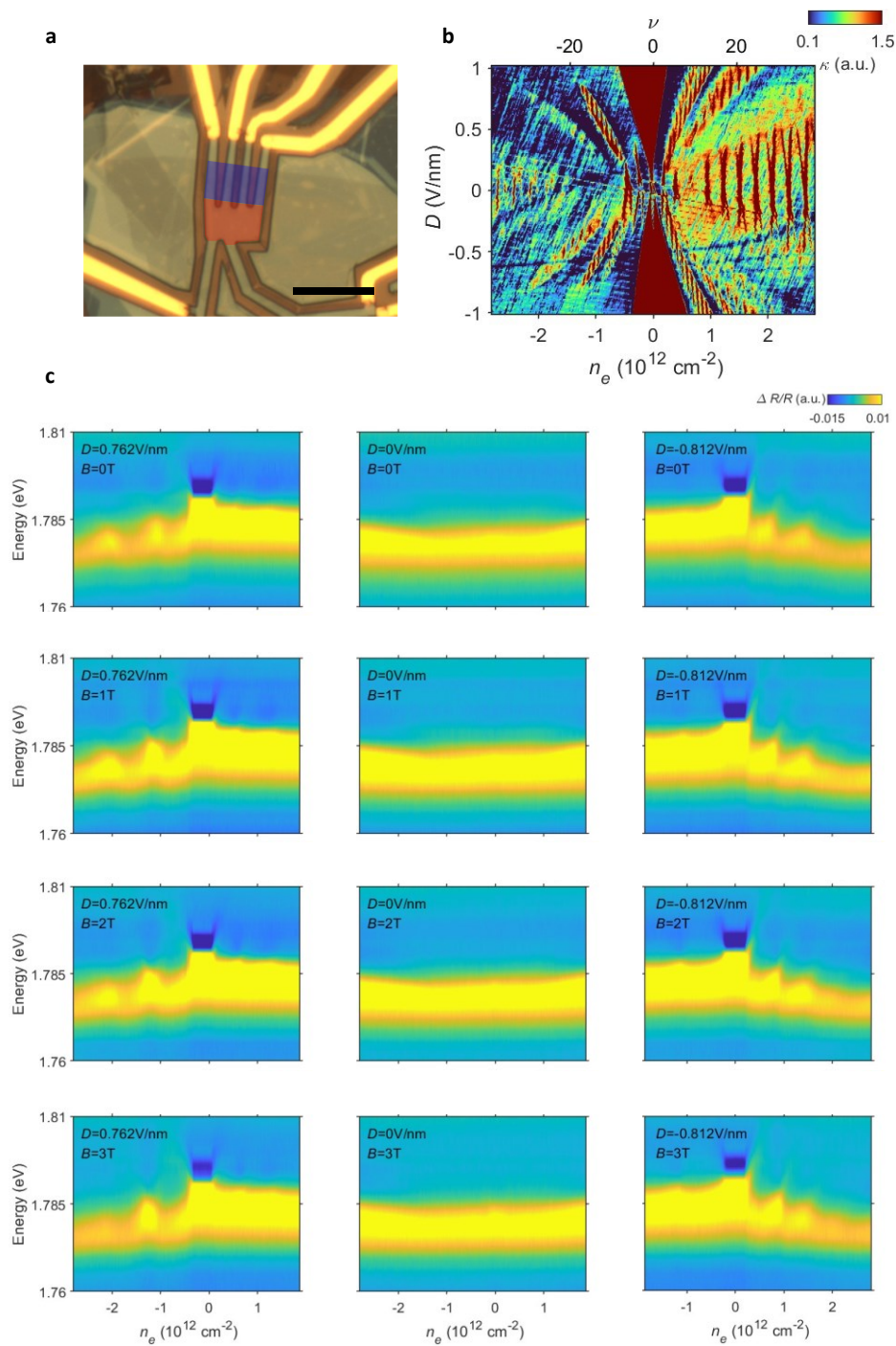

**Extended Data Figure 1. More results on device D1.** (**a**) Optical microscope image of RTG device D1. Scale bar: 10μm. It has two regions with (red) and without (blue) WSe$_2$, respectively, giving rise to two slightly offset patterns in compressibility (Fig. 1b and (**b**)). On the other hand, RC is not affected by such spatial inhomogeneity since it is a local measurement. (**b**) Displacement-field and carrier-density dependent incompressibility under magnetic field $B$=3T. Prominent incompressible peaks emerge from Landau gaps. (**c**) RC at representative displacement fields $D$=0.762, 0, and -

0.812V/nm under magnetic field $B$=0 to 3T. No features correspond to Landau gaps are observed. All measurements are performed at 3K.

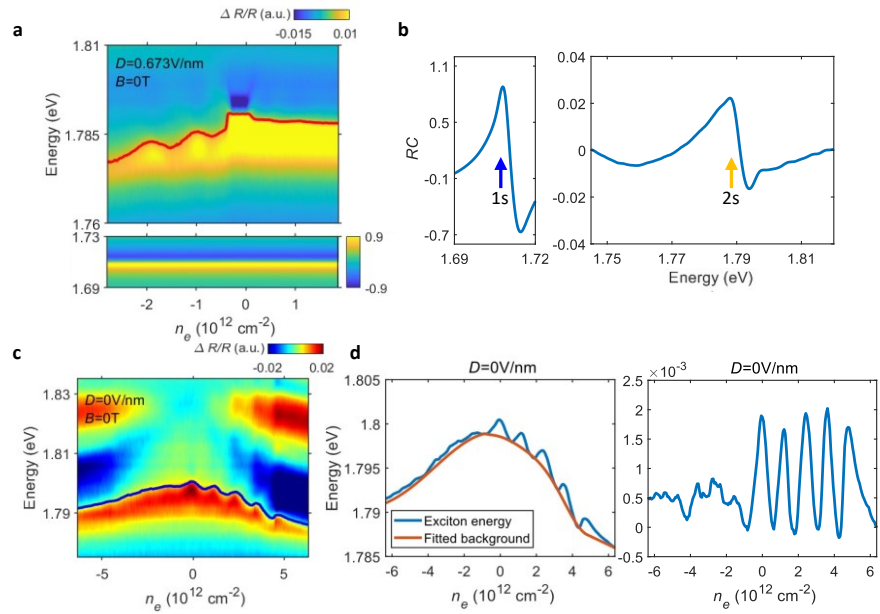

**Extended Data Figure 2. Extraction of 2s exciton energy**. (**a**) Upper panel: RC of device D1 at D=0.673V/nm with the extracted 2s exciton energy overlaid on top (red line). Lower panel: the RC near 1s exciton resonance, which remain largely unchanged over doping. (**b**) Representative RC spectrum of device D1 at $D$=0.673V/nm and $n_e$=0cm$^{-2}$. The 1s and 2s exciton resonances are marked by blue and yellow arrows, respectively. (**c**) RC of device D2 at $D$=0V/nm with the extracted 2s exciton energy overlaid on top (blue line). (**d**) Left: The extracted 2s exciton energy (blue) and the fitted background (orange). Right: background-subtracted 2s exciton energy showing clear peaks from the cascade features.

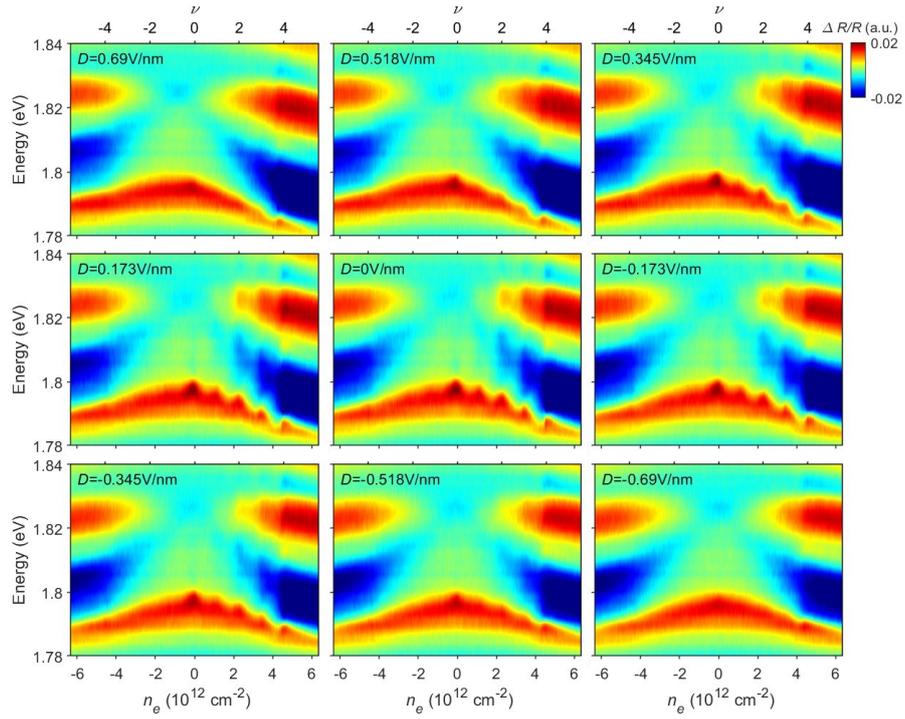

**Extended Data Figure 3. More results on device D2**. RC of MATTG device D2 at $D = \pm 0.69, \pm 0.518, \pm 0.345, \pm 0.173,$ and $0$ V/nm. The cascade features at integer filling become weaker at larger displacement field. All measurements are performed at 3K.

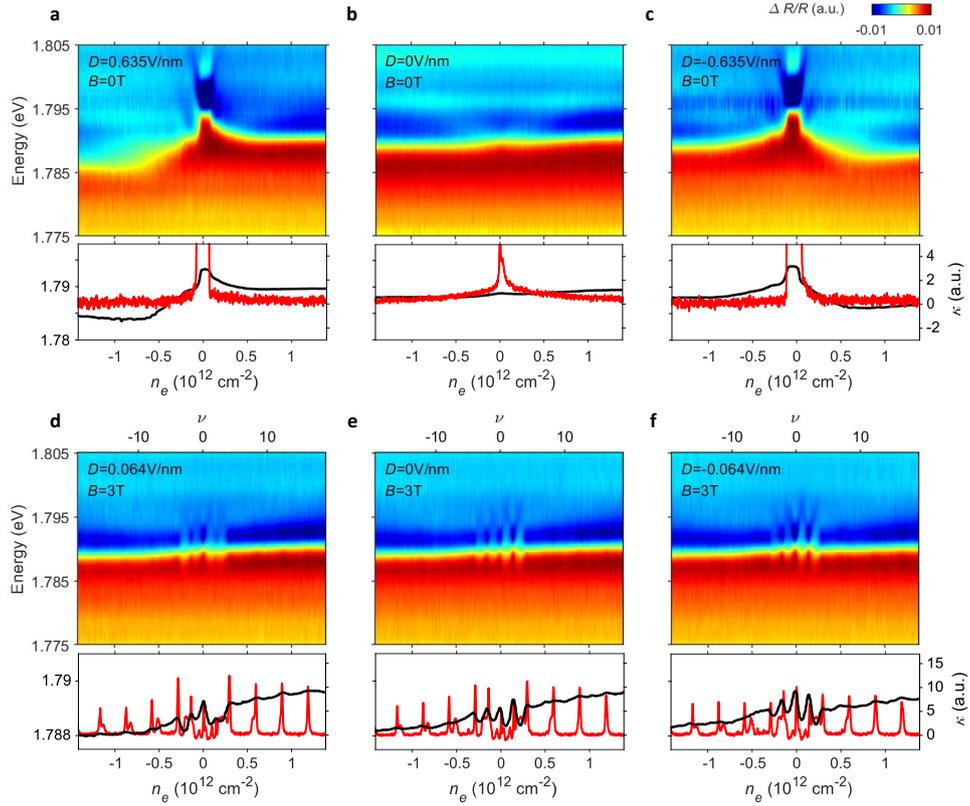

**Extended Data Figure 4. More results on device D3.** (**a**)-(**c**) Upper panel: RC of BBG device D3 at $B$=0T and $D$ = 0.635 (**a**), 0 (**b**), and -0.635V/nm (**c**). Lower panel: extracted 2s exciton energy (black) and comparison with incompressibility (red). (**d**)-(**f**) Upper panel: RC of device D3 at $B$=3T and $D$ = 0.064 (**d**), 0 (**e**), and -0.064V/nm (**f**). Lower panel: extracted 2s exciton energy (black) and comparison with incompressibility (red). The 2s exciton energy shows prominent oscillations within the zeroth Landau level.

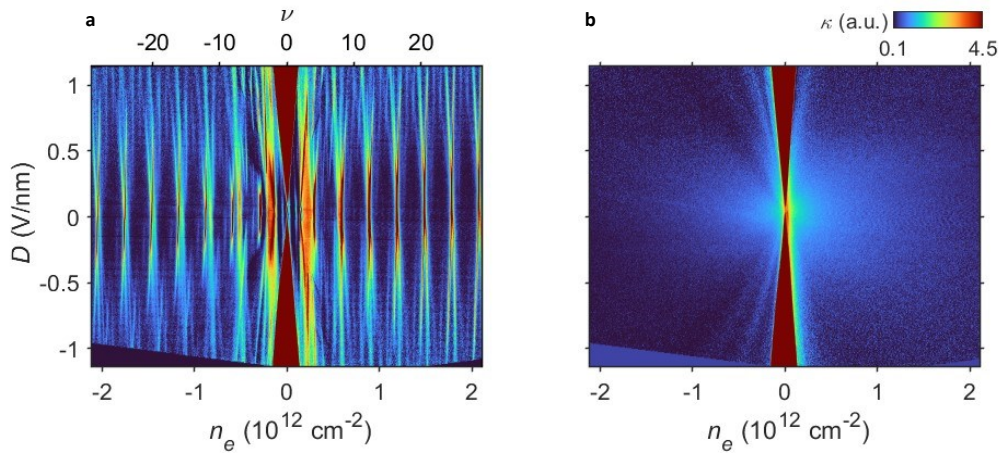

**Extended Data Figure 5. Incompressibility of device D3.** (**a**)(**b**) Incompressibility as a function of displacement field and carrier density at B= 3 (**a**) and 0T (**b**).

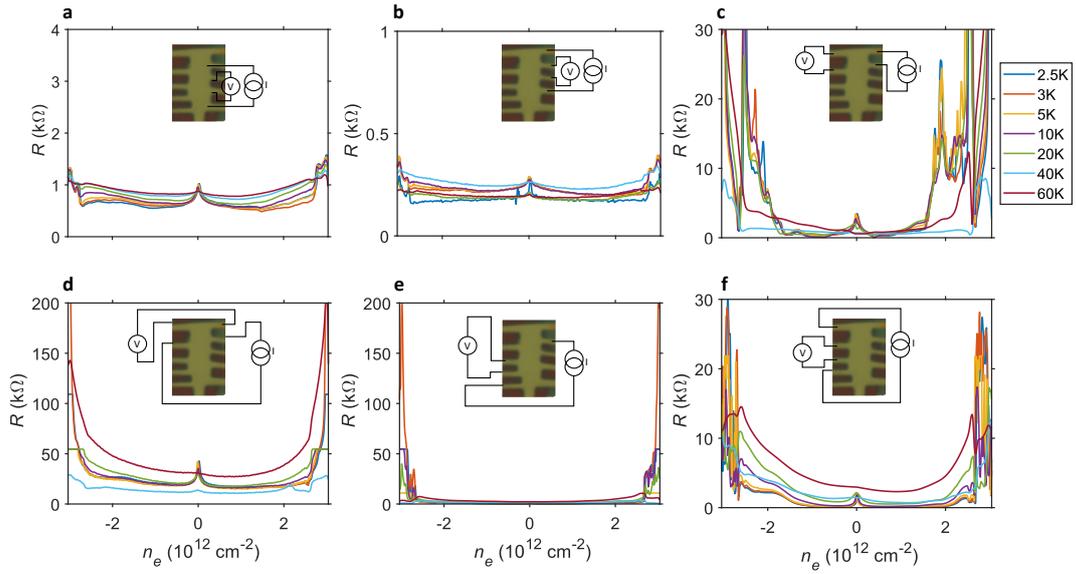

**Extended Data Figure 6. Longitudinal resistance of device D5**. (**a**)-(**f**) Temperature and carrier-density dependent longitudinal resistance of device D5 using different terminal configurations. The insulating feature at $v=3$ only appears in measurement configuration (**c**), consistent with the RC measurement.

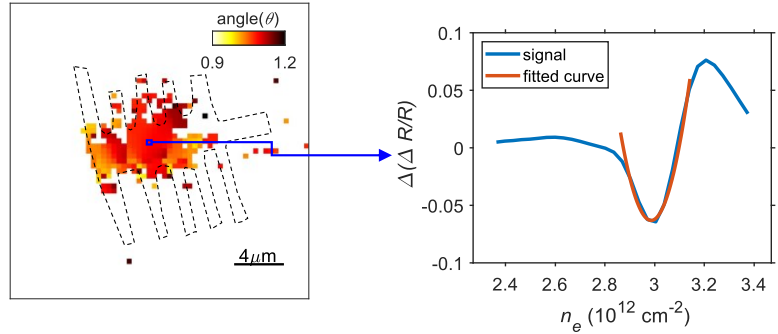

**Extended Data Figure 7. Wide-field imaging of device D5.** Left: spatial map of twist angle extracted from the $v = 4$ cascade feature. Right: carrier-density dependent differential reflection (blue) near $v = 4$ for the spatial spot marked by the blue box. The dip associated with cascade feature is fitted by a 2nd-order polynomial, as detailed in the methods. The extracted carrier density is $n_{v=4} \approx 2.95 \times 10^{12}$cm$^{-2}$, corresponding to 1.12° twist angle.

# Supplemental Information on Theoretical Methods: "Optical Imaging of Flavor Order in Flat Band Graphene"

Tian Xie,[1] Tobias M. Wolf,[2] Siyuan Xu,[1] Zhiyuan Cui,[1] Richen Xiong,[1] Yunbo Ou,[3] Patrick Hays,[3] Ludwig F Holleis,[1] Yi Guo,[1] Owen I Sheekey,[1] Caitlin Patterson,[1] Trevor Arp,[1] Kenji Watanabe,[4] Takashi Taniguchi,[5] Seth Ariel Tongay,[3] Andrea F Young,[1] Allan H. MacDonald,[2, *] and Chenhao Jin[1, †]

[1]*Department of Physics, University of California at Santa Barbara, Santa Barbara, CA, 93116, USA*

[2]*Department of Physics, University of Texas at Austin, Austin, TX, 78712, USA*

[3]*Materials Science and Engineering Program,
School of Engineering for Matter, Transport, and Energy,
Arizona State University, Tempe, Arizona 85287, USA*

[4]*Research Center for Electronic and Optical Materials,
National Institute for Materials Science, 1-1 Namiki, Tsukuba 305-0044, Japan*

[5]*Research Center for Materials Nanoarchitectonics,
National Institute for Materials Science, 1-1 Namiki, Tsukuba 305-0044, Japan*

Here we provide supplemental information summarizing our theoretical models for (i) the valence band exchange correction due to proximate polarizable layers, (ii) the polarizability of rhombohedral trilayer graphene at $B = 0$ obtained from band structure calculation, (iii) the polarizability of rhombohedral bilayer graphene at finite $B$ fields accounting for Landau quantization.

## EXCHANGE CORRECTION TO THE TMD VALENCE BAND

The dielectric environment surrounding a single-layer TMD, such as $WSe_2$, can strongly renormalize both the exciton binding energies and the single-particle band gap [1–3]. While for 1s excitons, the two effects are observed to be of the same order of magnitude (and thus cancel changes in the resonance energy), for $n$s excitons with $n \geq 2$, the effect of band gap renormalization is systematically observed to be dominant. The present work aims to highlight how this band gap shift in a TMD monolayer (the "sensing layer") can be used to witness electronic properties of a nearby target layer, and in particular highlights that this probe is sensitive to layer and flavor polarization.

The leading-order effect that changes the bandgap $E_{\text{gap}} \to E_{\text{gap}} + \delta E_{\text{gap}}$ is an exchange correction in the self-energy that strongly affects the valence band. The valence band shift $\delta E_{\text{vb}} = -\delta E_{\text{gap}}$ can be determined by examining the on-shell self-energy in the $G_0W$ approximation and evaluating the different contributions that arise in the analytical continuation. In our case, this leads to

$$\delta E_{\text{vb}} \simeq -\int \frac{d^2q}{(2\pi)^2} \left[ W_{11}(\boldsymbol{q}, \omega = |\epsilon_{\boldsymbol{q}}|) - V_{11}(\boldsymbol{q}) \right], \tag{1}$$

where $W_{11}$ is the retarded screened interaction between charge carriers within the insulating sensing layer in the presence of the target layer, and $V_{11}$ is the interaction without it. Note that $\epsilon_{\boldsymbol{q}}$ is the TMD dispersion near the valence band maximum. In the following sections, we will discuss how we model the screened interaction, and we evaluate it for examples relevant to the experiments in the main text. In the specific cases we are studying here, we will find that $\epsilon_{\boldsymbol{q}}$ is the smallest energy scale – which allows us to work in the static limit ($\omega \approx 0^+$).

## SCREENED INTERACTION WITHIN THE SENSING LAYER

The retarded screened intralayer-interaction $W_{11}(\boldsymbol{q},\omega)$ contains contributions induced by the charge response $\Pi_{22}(\boldsymbol{q},\omega)$ of the target layer. An explicit expression follows from $W(\boldsymbol{q},\omega) = \epsilon(\boldsymbol{q},\omega)^{-1} V(\boldsymbol{q})$, and the dielectric response is related to the bare interactions $V$ and the polarizability $\Pi$ through $\epsilon(\boldsymbol{q},\omega) = 1 - V(\boldsymbol{q})\Pi(\boldsymbol{q},\omega)$. We find

$$W_{11}(\boldsymbol{q},\omega) = V_{11}(\boldsymbol{q}) + \chi_{22}(\boldsymbol{q},\omega)V_{12}(\boldsymbol{q},\omega)^2, \qquad \chi_{22}(\boldsymbol{q},\omega) = \frac{\Pi_{22}(\boldsymbol{q},\omega)}{1 - V_{22}(\boldsymbol{q})\Pi_{22}(\boldsymbol{q},\omega)}, \qquad (2)$$

where $\chi_{22}$ is the charge response in the target layer and intra- and interlayer interactions are

$$V(\boldsymbol{q}) = \begin{pmatrix} V_{11}(\boldsymbol{q}) & V_{12}(\boldsymbol{q}) \\ V_{12}(\boldsymbol{q}) & V_{22}(\boldsymbol{q}) \end{pmatrix} = \frac{2\pi e^2}{q} \begin{pmatrix} 1/\epsilon_{11} & e^{-qd}/\epsilon_{12} \\ e^{-qd}/\epsilon_{12} & 1/\epsilon_{22} \end{pmatrix}. \qquad (3)$$

The interlayer distance $d$ is about $d \sim 2a_G$ without a spacer and about $d \sim 20a_G$ with a spacer. For the dielectric constants, for simplicity, we assume $\epsilon_{ij} \simeq \epsilon_r \simeq 5$ without affecting our results significantly. It is worth highlighting, that here we assume that the sensing layer interacts with a single target layer. If needed, we could straightforwardly extend this model to account for additional layer degrees of freedom in $V$ and $\Pi$, but we will instead focus on qualitative aspects unrelated to layer polarization in the target system.

In the random-phase approximation (RPA), the polarizability $\Pi_{22}(\boldsymbol{q},\omega)$ in the target layer is just the bare polarization bubble diagram. In terms of the non-interacting single-particle eigenbasis, we have

$$\Pi_{22}(\boldsymbol{q},\omega) = \frac{1}{A_{\text{uc}}} \sum_{n,m,\alpha,\boldsymbol{k}} \frac{f(\varepsilon_{n\alpha}(\boldsymbol{k})) - f(\varepsilon_{m\alpha}(\boldsymbol{k}+\boldsymbol{q}))}{\varepsilon_{n\alpha}(\boldsymbol{k}) - \varepsilon_{m\alpha}(\boldsymbol{k}+\boldsymbol{q}) + \omega + i\delta} \times |\langle \psi_{n\alpha}(\boldsymbol{k})|\psi_{m\alpha}(\boldsymbol{k}+\boldsymbol{q})\rangle|^2, \qquad (4)$$

where $\varepsilon_{n\alpha}(\boldsymbol{k})$ is the energy band $n$ for orbital index $\alpha$, and $|\psi_{m\alpha}(\boldsymbol{k})\rangle$ the corresponding wavefunctions, and $A_{\text{uc}}$ is the unit cell area. The overlap matrix element is often referred to as form factor. Similarly, in presence of a static homogeneous transverse magnetic field $\boldsymbol{B} = B\hat{z}$, the polarizability in terms of the Landau level spectrum is [4]

$$\Pi_{22}(\boldsymbol{q},\omega) = \frac{1}{2\pi\ell^2} \sum_{n\alpha,n'\alpha'} \frac{\Theta(\epsilon_F - \varepsilon_{n\alpha}) - \Theta(\epsilon_F - \varepsilon_{n\alpha})}{\varepsilon_{n\alpha} - \varepsilon_{n\alpha} + \omega + i\delta} \left|\mathcal{F}_{n\alpha,n'\alpha'}(\boldsymbol{q})\right|^2, \qquad (5)$$

where $\varepsilon_{n\alpha}$ are the Landau levels for integer $n$ and orbital $\alpha$, and $\mathcal{F}_{n\alpha,n'\alpha'}(\boldsymbol{q})$ are multiorbital magnetic form factors, which are given by orbital superpositions of the magnetic form factors of the two-dimensional electron gas.

In Fig. 1, we show the static polarizabilities for flavor-polarized states in rhombohedral trilayer graphene in a large displacement field, and in Fig. 2 we show the static polarizability of bernal bilayer graphene in a strong magnetic field at filling factors corresponding to the nearly-orbital-degenerate lowest two Landau levels. In both cases, we find that spontaneous polarization of electronic states in the graphene material should lead to significant band gap renormalization in a nearby WSe$_2$ sensing layer, in agreement with experiment observations.

## RHOMBOHEDRALLY-STACKED MULTILAYER GRAPHENE

In what follows, we will briefly review the band structure model and Landau quantization for the rhombohedral multilayer graphene that we employ to evaluate polarizabilities in Eqs. (4) and (5). In giving explicit expressions, we will focus on the trilayer case – the bilayer case requires minor changes to the basis and to the tight-binding parameters.

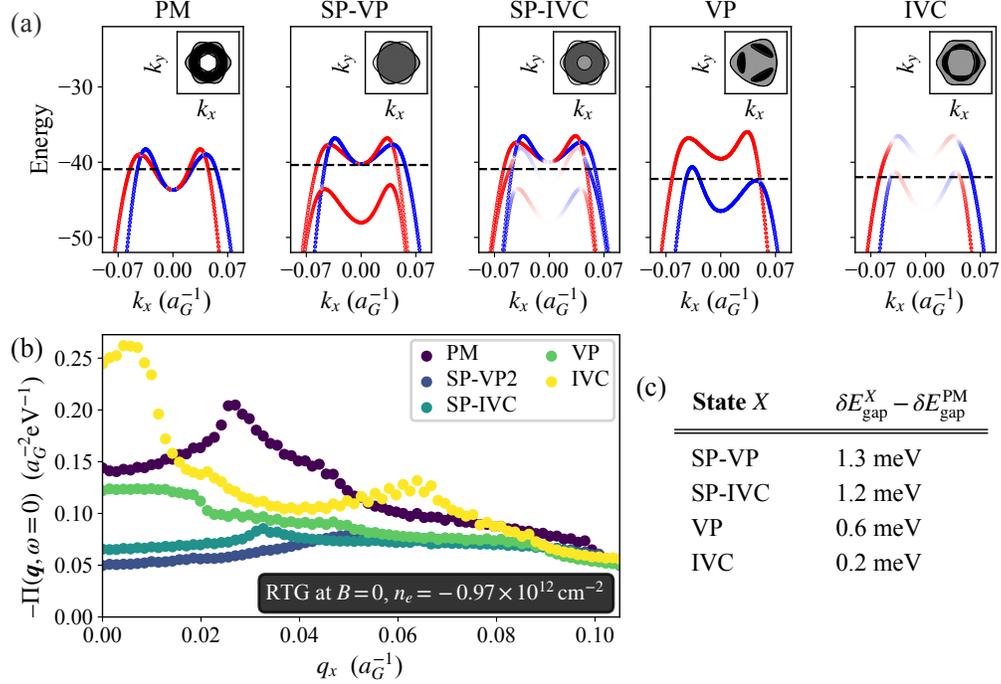

Figure 1. Static polarizability of spin–valley polarized states in rhombohedral trilayer graphene (RTG) at large displacement field $U_D = 30$ meV and fixed electron density $n_e = -0.97 \times 10^{12}$ cm$^{-2}$. (a) Mean-field bandstructures of the paramagnetic/symmetric state (PM), the spin and valley-polarized state (SP-VP), the spin-polarized intervalley-coherent (SP-IVC), valley-polarized (VP), and intervalley-coherent (IVC) state, where band colors indicate the valley expectation values (red: $K$, blue: $K'$). Insets illustrate the corresponding Fermi sea and surfaces (hue: hole occupation number). (b) Static polarizability vs $q_x$ momentum for all candidate states, with stark differences that are associated with the phase space for particle–hole excitations on the Fermi sea at given $q$. (c) Predicted bandgap renormalization in a nearby monolayer of WSe$_2$ when the system undergoes symmetry breaking from PM to symmetry-broken state, obtained by evaluating Eq. (1) using the static polarizability shown in panel (b). Note that the candidate states were computed using self-consistent Fock mean-field theory using dielectric permittivity $\epsilon_r = 5.5$ and screening gate distance $d_{\text{gate}} = 50$ nm.

*Continuum model.* Rhombohedral trilayer graphene (RTG) has a multi-layered triangular lattice structure with lattice constant $a = 2.46$ Å and 6 atoms per unit cell: three layers (separated by $d = 3.4$ Å), each with two sublattices. Neighboring layers and next-neighboring layers are mutually AB-stacked. We employ the commonly-used and well-known continuum model for the low-energy dispersion of $\pi$-electrons with spin $s = \pm 1/2$ near each nonequivalent valley $\tau K$ with $\tau = \pm 1$ at Brillouin zone corners [5, 6]. We label the four spin–valley combinations as flavor index $\alpha = (s, \tau)$. Ordering the basis as A1, B1, A2, B2, . . . , the resulting continuum Hamiltonian per flavor is [5]

$$h(\boldsymbol{k}) = \begin{bmatrix} t(\boldsymbol{k}) + U_1 & t_{12}(\boldsymbol{k}) & t_{13} \\ t_{12}^\dagger(\boldsymbol{k}) & t(\boldsymbol{k}) + U_2 & t_{12}(\boldsymbol{k}) \\ t_{13}^\dagger & t_{12}^\dagger(\boldsymbol{k}) & t(\boldsymbol{k}) + U_3 \end{bmatrix}_{6 \times 6}, \qquad (6)$$

where $\boldsymbol{k} = (k_x, k_y)$ is the Bloch momentum measured w.r.t. valley $\tau$. The model contains intralayer hopping ($t$), nearest-layer hopping ($t_{12}$), and next-nearest-layer hopping ($t_{13}$):

$$t(\boldsymbol{k}) = \begin{bmatrix} 0 & v_0 \pi^\dagger \\ v_0 \pi & 0 \end{bmatrix}, \quad t_{12}(\boldsymbol{k}) = \begin{bmatrix} -v_4 \pi^\dagger & v_3 \pi \\ \gamma_1 & -v_4 \pi^\dagger \end{bmatrix}, \quad t_{13}(\boldsymbol{k}) = \begin{bmatrix} 0 & \gamma_2/2 \\ 0 & 0 \end{bmatrix}, \qquad (7)$$

where $\pi = \tau k_x + i k_y$ is the linear momentum, and $\gamma_i$ ($i = 0, \ldots, 4$) are tight-binding parameters with corresponding velocity parameters $v_i = (\sqrt{3}/2) a \gamma_i / \hbar$. We use the model parameters listed in Table I, which are chosen to match quantum oscillation frequency signatures in Ref. [7]. We include different layer potentials (through the $U_i$-terms) induced by top and bottom gate, to coreclty account for the electric displacement field $D$ and the electronic density $n_e$. We note that intrinsic Ising-type spin–orbit coupling (SOC) $\lambda$ is negligible (about 10–50 $\mu$eV), but proximity to transition metal dichalcogenides (e.g., WSe$_2$ without spacer) can drastically enhance it in the nearest graphene layer (up to $\lambda \sim 0.800$ meV). The impact of proximity-induced SOC on the electronic bands is highly sensitive to the sign of the displacement field $D$. In this work, we neglect SOC to focus on qualitative effects and leave the analysis of SOC for future studies.

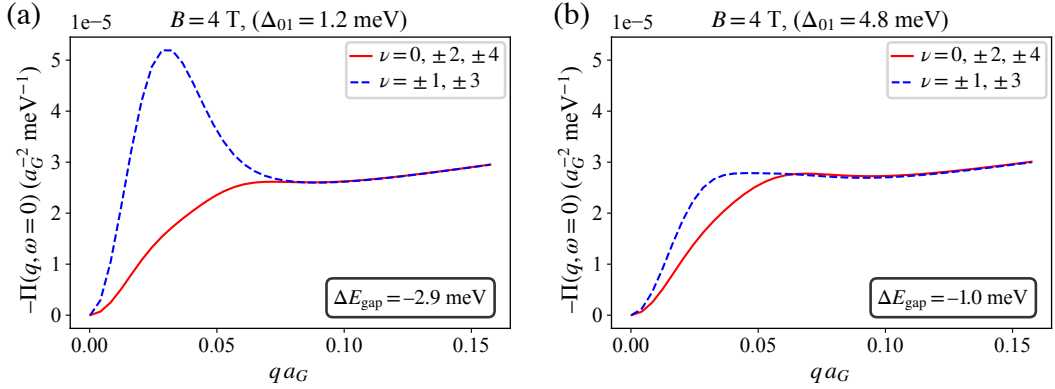

Figure 2. Static polarizability of the Landau-quantized electronic spectrum of bernal bilayer graphene at magnetic field $B = 4$ T and vanishing displacement field $U_D = 0$ at filling factors $\nu = -4, \ldots, 4$. Due to interactions, the filling of the lowest Landau levels occurs flavor by flavor. The polarizability shows a strong even/odd dependence, due to additional interorbital transitions in each flavor sector between Landau level 0 and Landau level 1 that are separated by a gap $\Delta_{01}$. We show two different cases for $\Delta_{01}$ in panel (a) and panel (b), respectively, and indicate the resulting difference in the proximity-induced WSe$_2$ bandshift between even and odd filling, $\Delta E_{\text{gap}} = \delta E_{\text{gap}}^{\text{odd}} - \delta E_{\text{gap}}^{\text{even}}$. In the results shown here, we imposed flavor polarization without performing self-consistent Hartree-Fock calculations.

Table I. Tight-binding parameters (in eV) for rhombohedral trilayer graphene, see also Refs. [5–7].

| $\gamma_0$ | $\gamma_1$ | $\gamma_2$ | $\gamma_3$ | $\gamma_4$ | $U$ | $\Delta$ | $\delta$ |
|---|---|---|---|---|---|---|---|
| 3.160 | 0.380 | −0.015 | −0.290 | 0.141 | 0.030 | −0.0023 | −0.0105 |

*Landau quantization.* Transverse magnetic fields quantize the electronic spectrum into Landau levels, described by a minimal coupling $\boldsymbol{p} \mapsto \boldsymbol{\pi} = \boldsymbol{p} - e\boldsymbol{A}$ in Eq. (6). The commutation relation of $\pi = \pi_x + i\pi_y$ with $\pi^\dagger$ implies ladder operators $\hat{a}$ and $\hat{a}^\dagger$ with $[\hat{a}, \hat{a}^\dagger] = 1$ such that $\pi_x = \frac{\hbar}{\sqrt{2}l_B}(\hat{a}^\dagger + \hat{a})$, $\pi_y = \frac{\hbar}{i\sqrt{2}l_B}(\hat{a}^\dagger - \hat{a})$. We defined the magnetic length $\ell_B = \sqrt{\hbar/eB}$. Each Landau level has degeneracy $\Phi/\phi_0$, where $\Phi = BA$ is the flux and $\phi_0 = h/e$ a flux quantum, and thus accommodates the electronic density $n_B = 1/(2\pi\ell_B^2)$. Writing the Hamiltonian in the eigenbasis of $\hat{N} = \hat{a}^\dagger \hat{a}$ allows to obtain the electronic spectrum (numerically and in simplified cases analytically), see Fig. 3 for the case of bernal bilayer graphene (BBG).

*Flavor magnetism.* Experiments studying top/back-gated rhombohedral multilayers demonstrate that correlated states such as spin–valley magnetism, superconductivity and fractional anomalous quantum Hall states are ubiquitous in these materials [7–13]. The simple Coulomb potential commonly used to approximate effective intralayer interactions is $V_{\boldsymbol{q}} = (2\pi k_e/\epsilon_r) \tanh(|\boldsymbol{q}|d_{\text{gate}})/|\boldsymbol{q}|$, for metallic gates at distance $d_{\text{gate}} \in [30, 60]$ nm, effective relative permittivity $\epsilon_r \in [7, 20]$, and Coulomb constant $k_e = 1.44$ eV nm. The spin–valley magnetism can (to a limited degree) be understood from a mean-field perspective: the energy gain in exchange energy by sequentially occupying individual flavors can outweigh the cost associated with the kinetic band energy – especially if the band dispersion is flat/the density of states is large. In Fig. 4, we illustrating this effect by showing the total energy of different spin–valley polarized states obtained from self-consistent Hartree-Fock mean-field calculations for the case of bernal bilayer graphene.

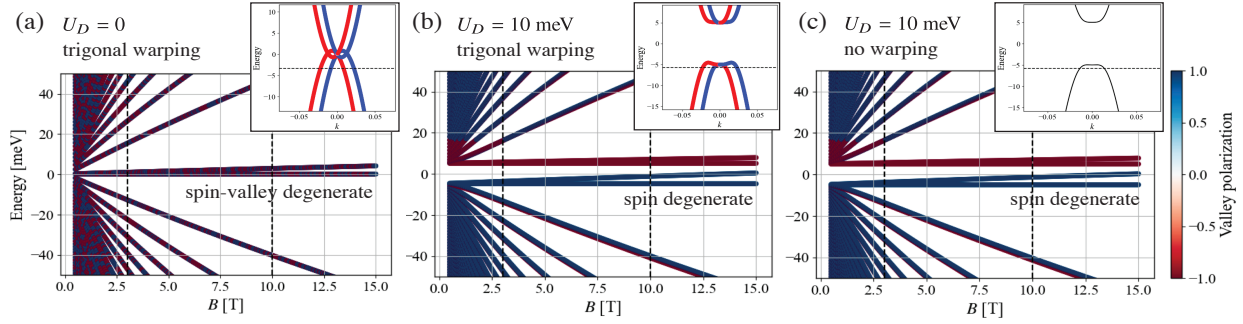

Figure 3. Landau level spectra for bernal bilayer graphene (BBG) (sketched in inset) at large displacement fields. Insets show band structures at zero field $B = 0$. (a) Landau levels at zero displacement field, such that the lowest two (anomalous) landau levels are spin and valley degenerate. (b) Landau levels at finite displacement field, lifting the valley degeneracy in the anomalous Landau levels. (c) Landau levels when trigonal warping is neglected, illustrating that trigonal warping is negligible at large magnetic fields.

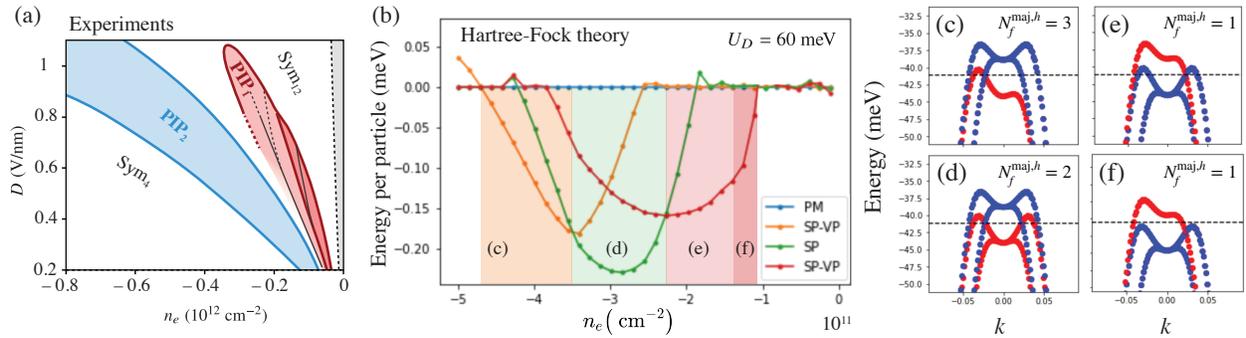

Figure 4. Phase diagram of bernal bilayer graphene with partially and fully polarized magnetic phases (PIP, half-metal, quarter-metal). (a) Sketch of the phase diagram as observed experimentally in Ref. [11], with symmetric/paramagnetic phases (Sym$_4$, Sym$_{12}$) and partially polarized phases (PIP$_1$, PIP$_2$). (b) Hartree-Fock ground state energies of different types of paramagnetic and (c–f) spin–valley polarized states (PM, SP-VP, SP). The magnetic states differ by the number $N_f^{\mathrm{maj},h}$ of majorly hole-occupied flavors. (c–f) Examples for mean-field band structures of the magnetic states. Here, for simplicity, we do not include intervalley coherent states in the analysis. In panels (b–f), we used $U_D = 62$ meV, $\epsilon_r = 4.4$, $d_{\mathrm{gate}} = 40$ nm.

---

* macdpc@physics.utexas.edu
† jinchenhao@ucsb.edu